\documentstyle[aps, prl, epsf, floats]{revtex}

\begin{document}

\onecolumn

\newcommand{\mb}[1]{\mbox{\boldmath $#1$}}
\newenvironment{equationarray}
  {\arraycolsep 0.14 em \begin{eqnarray}}{\end{eqnarray}}

\makeatletter
  \def\gsim{\compoundrel>\over\sim}
  \def\lsim{\compoundrel<\over\sim}
  \def\compoundrel#1\over#2%
  {\mathpalette\compoundreL{{#1}\over{#2}}}
  \def\compoundreL#1#2{\compoundREL#1#2}
  \def\compoundREL#1#2\over#3%
  {\mathrel{\vcenter{\hbox{$\m@th \buildrel{#1#2}\over{#1#3}$}}}}
\makeatother

\twocolumn[\hsize\textwidth\columnwidth\hsize\csname
  @twocolumnfalse\endcsname
\draft

\title{Gravitational waves from relativistic rotational core collapse}
\author{Harald Dimmelmeier, Jos\'e A.~Font and Ewald M\"uller}
\address{Max-Planck-Institut f\"ur Astrophysik,
Karl-Schwarzschild-Str.~1, D-85740 Garching, Germany.}
\date{\today}
\maketitle

\begin{abstract}
  We present results from simulations of axisymmetric relativistic
  rotational core collapse. The general relativistic hydrodynamic
  equations are formulated in flux-conservative form and solved using
  a high-resolution shock-capturing scheme. The Einstein equations are
  approximated with a conformally flat 3-metric. We use the quadrupole
  formula to extract waveforms of the gravitational radiation
  emitted during the collapse. A comparison of our results with those
  of Newtonian simulations shows that the wave amplitudes
  agree within 30\%. Surprisingly, in some cases, relativistic
  effects actually diminish the amplitude of the gravitational wave
  signal. We further find that the parameter range of models suffering
  multiple coherent bounces due to centrifugal forces is considerably
  smaller than in Newtonian simulations.
\end{abstract}


\vskip2pc]

Since more than two decades astrophysicists struggle to compute the
gravitational wave signal produced by rotational core collapse
supernovae (see e.g.~\cite{mueller_95}). Besides the coalescence of
compact black hole and/or neutron star binaries, these events are
considered as one of the most promising (burst) sources to detect
gravitational waves~\cite{thorne_97}. Theoretical predictions have
been, and still are, hampered by two major reasons: Firstly, one does
not know the rotational state of the core prior to its gravitational
collapse, as multi-dimensional evolutionary calculations of rotating 
stars have not yet been performed. Up to now only quasi-spherical
stellar pre-collapse models with a phenomenological description of
mixing and angular momentum transport are
available~\cite{heger_00}. Thus, predictions from rotational core
collapse must rely upon extensive parameter studies, i.e., a large set
of simulations. Secondly, reliable collapse simulations require a
relativistic treatment of the dynamics, as both high densities and
velocities together with strong gravitational fields are encountered,
as well as the incorporation of a realistic equation of state (EoS)
and neutrino transport. All previous studies have either neglected or
approximated one or all of these requirements. In addition, most
simulations assumed axisymmetry, the fully three-dimensional case
still remaining essentially unexplored.

Concerning the gravitational wave signal the two most advanced studies
of axisymmetric rotational core collapse to date have either
investigated a few Newtonian models with a realistic EoS and simplified
neutrino treatment~\cite{moenchmeyer_91}, or a large set of Newtonian
models with a simplified EoS and no neutrino
transport~\cite{zwerger_97}. These simulations predict that
nonspherical core collapse supernovae emit at most
$ \sim 10^{-6} M_\odot c^2 $ ($ M_\odot = 1.99 \times 10^{33} {\rm\ g} $
is the sun's mass, and $ c $ is the speed of light in vacuum) in the
form of gravitational radiation, with a maximum quadrupole wave
amplitude $ A_{20}^{\rm E2} \sim 10^3 {\rm\ cm} $. For a source at a
distance $ R $ this corresponds to a maximum dimensionless strain
$ h \sim 10^{-23} \cdot (10 {\rm\ Mpc} / R) $. Similar maximum signal
strengths have been found in non-axisymmetric Newtonian simulations
also using a simplified EoS and neglecting neutrino
transport~\cite{rampp_98}.

The present work is a further step to improve the reliability of the
prediction of the gravitational wave signal from rotational core
collapse. To this end we have performed the first study (to our
knowledge) in the relativistic framework, extending the axisymmetric 
Newtonian core collapse simulations of Zwerger and 
M\"uller~\cite{zwerger_97}. 

In order to examine a large set of initial conditions we approximate
the collapsing core by a rotating polytrope and use a simplified
analytic EoS~\cite{janka_93}. We neglect all transport
effects but include relativistic dynamics in a time-dependent
spacetime. The pressure $ P $ consists of a polytropic part
$ P_{\rm p} $ and a thermal part $ P_{\rm th} $,
\begin{equation}
  P (\rho, \varepsilon) = P_{\rm p} + P_{\rm th} =
  K \rho^{\gamma_{\rm p}} + \rho \varepsilon_{\rm th}
  (\gamma_{\rm th} - 1),
  \label{eq:eos}
\end{equation}
where $\rho$ is the rest-mass density, and 
$\varepsilon = \varepsilon_{\rm p} + \varepsilon_{\rm th} $ is the
specific internal energy which consists of a polytropic and a thermal
contribution. The polytropic constant $K$ and the polytropic index
$ \gamma_{\rm p} $ change discontinuously at nuclear matter density,
$ \rho_{\rm nuc} = 2.0 \times 10^{14} {\rm\ g\ cm}^{-3} $, but are
otherwise assumed to be independent of $\rho$ elsewhere.
$ P_{\rm th} $ mimics the thermal pressure in the matter heated up by
the shock. Writing $ P $ as a sum of these two terms prevents matter
elements from following the same $ P $--$ \rho $-history during infall
and expansion after bounce (for more details see~\cite{janka_93}). In
all models we set $ \gamma_{\rm th} = 1.5 $, which corresponds to a
mixture of relativistic and nonrelativistic gas.

Initial configurations are rotating relativistic stellar cores in
equilibrium computed by means of Hachisu's self-consistent field
method~\cite{komatsu_89_a}. All initial models are 4/3-polytropes,
i.e., $ P = K \rho^{4/3} $, with a central density of
$ 1.0 \times 10^{10} {\rm\ g\ cm}^{-3} $. They rotate according to the
differential rotation law specified in~\cite{komatsu_89_a}, in which a
parameter $ A $ (a length scale) determines the angular momentum
distribution. For large values of $ A $ (compared to the size of the
initial model) one obtains almost rigidly rotating configurations,
while small values of $ A $ correspond to strongly differentially
rotating ones. This rotation law is the relativistic extension of the
Newtonian one in~\cite{eriguchi_84}. The second parameter of the
initial model is the ratio of the amount of rotational energy to the
modulus of the gravitational potential energy,
$ \beta = E_{\rm rot} / |E_{\rm pot}| $. The collapse is initiated by
suddenly reducing $ \gamma_{\rm p} $ from its initial value 4/3 to
some prescribed value in the range $ [1.28, 1.325] $. In all
simulations we assume both axial and equatorial plane symmetry, and do
not consider any physical angular momentum transport.

The hydrodynamic evolution of a relativistic fluid in an arbitrary
spacetime endowed with a metric $ g_{\mu \nu} $ is governed by the local
conservation laws of the current density $ J^\mu = \rho u^\mu $ and
the (perfect fluid) stress-energy $ T^{\mu \nu} = \rho h u^\mu u^\nu + 
P g^{\mu \nu} $, where $u^\mu$ is the 4-velocity and $h=1+\varepsilon +
P/\rho$ is the specific enthalpy. The conservation laws are the
continuity equation $ \nabla_\mu J^\mu = 0 $ and the Bianchi identities
$ \nabla_\mu T^{\mu \nu} = 0 $, respectively. To derive the hydrodynamic
equations~\cite{banyuls_97} we introduce a set of conserved variables
$ D = \rho W $ (rest-mass density), $ S_i = \rho h W^2 v_i $ (momentum
density) and $ E = \rho h W^2 - P $ (total energy density),
$ W = \alpha u^0 $ being the relativistic Lorentz factor, in terms of
the primitive variables $ (\rho, v_i, \varepsilon) $, where $ v_i $ is
the 3-velocity. The resulting equations can be written as a
first-order flux-conservative hyperbolic system of conservation laws.

The time-dependent metric is evolved using the $ 3 + 1 $ splitting of
spacetime. In this formalism the Einstein equations are written as a 
system of evolution and constraint equations for the 3-metric
$ \gamma_{ij} $ and the extrinsic curvature $ K^{ij} $. Unfortunately,
achieving long-term stable evolution using the $ 3 + 1 $ system (or
improved  reformulations) is still an open issue
(see~\cite{alcubierre_00} and references therein). Therefore,
following~\cite{CFC} we approximate the spacetime geometry by assuming
that the 3-metric is conformally flat, i.e.,
$ \gamma_{ij} = \phi^4 \hat{\gamma}_{ij} $ where $ \phi $ is a
conformal factor depending on the coordinates $ x^\mu $, and where
$ \hat{\gamma}_{ij} $ is the flat 3-metric. In this  approximation,
the Einstein equations transform into a system of 5 coupled nonlinear
elliptic equations for the conformal factor $ \phi $, the lapse
function $ \alpha $ and the shift vector $ \beta^i $:
\begin{equationarray}
  \hat{\Delta} \phi          & = & - 2 \pi \phi^5 (\rho h W^2 - P) -
  \frac {\phi^5 K_{ij} K^{ij}} 8,
  \label{eq:metric_eq_phi} \\
  \hat{\Delta} (\alpha \phi) & = & 2 \pi \alpha \phi^5 (\rho h
  (3 W^2 - 2) + 5 P) + \frac {7 \alpha \phi^5 K_{ij} K^{ij}} 8,
  \label{eq:metric_eq_alpha} \\
  \hat{\Delta} \beta^i       & = & 16 \pi \alpha \phi^4 S^i + 2 K^{ij}
    \hat{\nabla}_j \left( \frac {\alpha} {\phi^6} \right) - \frac 1 {3}
    \hat{\nabla}^i \hat{\nabla}_k \beta^k,
  \label{eq:metric_eq_beta}
\end{equationarray}%
with $ \hat{\Delta} $ and $ \hat{\nabla} $ being the flat space
Laplace and Nabla operators, respectively.

The above approximation is exact in spherical symmetry. As our focus
is to investigate rotational core collapse, where the matter 
distribution does not deviate too much from spherical symmetry
(contrary to binary neutron star mergers~\cite{CFC}),
and as the effect of the gravitational radiation reaction force onto
the hydrodynamics is negligible, the conformally flat condition (CFC)
appears as an adequate approximation for our purpose. The accuracy of
the CFC approximation has been considered by Cook
{\it et al.}~\cite{cook_96}, who find remarkably good results
for rapidly rotating relativistic stars in equilibrium, with typical
errors for different variables smaller than 5\%. However, the quality
of the CFC approximation degrades in the case of extremely
relativistic nonspherical configurations, e.g., rigidly rotating
infinitesimally thin disks of dust~\cite{kley_99}.

As off-diagonal elements of $ \gamma_{ij} $ are set to zero in the CFC
approximation, the gravitational wave content of the spacetime (the
transverse part of the metric tensor) is eliminated. Therefore, to
compute the gravitational wave signal we use the Einstein quadrupole
formula in a post-processing step.

In order to exploit the hyperbolic and conservative character of the
hydrodynamic equations, we have implemented in our code a Godunov-type
flux-conservative finite volume scheme based upon the solution of
approximate Riemann problems at each cell interface (see,
e.g.~\cite{font_00}). On each time slice, the hydrodynamic data are
represented by cell averages of the conserved quantities $ D $, $ S_i $
and $ E $. These are propagated to the next time level by a conservative
Runge--Kutta algorithm, which involves the numerical fluxes and
sources. The metric
equations~(\ref{eq:metric_eq_phi}--\ref{eq:metric_eq_beta}) are
discretized on a 2-dimensional grid with a 9-point stencil. A
Newton--Raphson iteration scheme is used to obtain the solution for
the 5 metric functions. For each metric computation, we have to solve
a linear system of equations involving the (sparse) Jacobian matrix of
the discretized equations.

We have performed a comprehensive number of tests of our code. Firstly,
we have studied its ability to keep rotating neutron stars in equilibrium. 
To that end we have solved the relativistic hydrostatic equilibrium equations
for rotating axisymmetric polytropes having different central densities,
angular momentum profiles, and rotation velocities up to the mass-shedding
limit. These initial data were evolved both on a static background
metric and in a fully dynamic spacetime, always finding stability of the
equilibrium configurations over many rotation periods. Secondly,
we have compared the performance of our Eulerian code with a
May--White-type Lagrangian artificial viscosity
code~\cite{may_66} in spherical symmetry. When evolving identical
initial collapse models with both codes we observe agreement in the
evolution of fluid and metric variables within a few percent.

In order to pin down the effects of general relativity on rotational
core collapse, we have compared relativistic (R) and Newtonian (N)
simulations for a large set of initial models. 
The collapse simulations have been performed in spherical coordinates
using a logarithmic radial grid with 200 zones and an equidistant
angular grid with 30 zones. Test runs with more angular zones
yield similar results. During a typical relativistic simulation the
metric is calculated every 100th (10th) time step for a central
density smaller (higher) than $ 0.1 \rho_{\rm nuc} $, and extrapolated
in between. Comparison with a run
where the metric was calculated every 10th (or every) time step showed
no significant differences. The total mass $ \int dV \sqrt{\gamma} D $
is conserved to better than $ 10^{-3} $ in both relativistic and
Newtonian rotational collapse. In spherical symmetry, the mass is
conserved up to machine precision in both the Newtonian case, and the
relativistic case for a static background metric.

\begin{table}
  \caption{Model parameters. Note that model~B never reaches
    supranuclear densities.}
  \begin{tabular}{cccccc}
    Model & $ A $ & $ \beta_{\rm R} $ & $ \beta_{\rm N} $ &
            $ \gamma_{\rm p} $ & $ \gamma_{\rm p} $ \\
          &  [$ 10^8 $ cm] & [\%] & [\%] & $ (\rho \le \rho_{\rm nuc}) $ &
            $ (\rho > \rho_{\rm nuc}) $ \\
    \hline
    A     &  0.5 & 0.47 & 0.48 & 1.300 & 5/3 \\
    B     &  1.0 & 1.82 & 1.81 & 1.325 & --- \\
  \end{tabular}
  \label{table:params}
\end{table}

\begin{figure}[t]
  \def\epsfsize#1#2{0.48#1}
  \centerline{\epsfbox{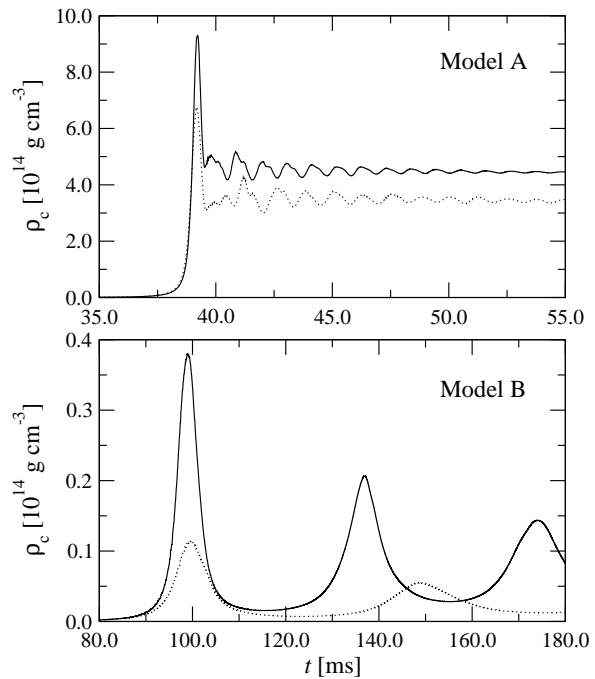}}
  \caption{Central density vs.\ time for model~A (upper panel) and B
    (lower panel); relativistic runs are denoted by solid lines,
    Newtonian runs by dotted lines.}
  \label{fig:rho_c_t}
\end{figure}

Here we present the results from two representative models
(Table~\ref{table:params}). Note that the initial data slightly differ
with respect to the value of $ \beta $ between the relativistic and
Newtonian initial model. 
The time-evolution of the central density is plotted in
Fig.~\ref{fig:rho_c_t}. Model~A is a strongly differentially rotating
core which collapses rapidly ($ t_{\rm bounce} = 38 {\rm\ ms} $) due to
the large reduction of $ \gamma_{\rm p} $ from its initial value 4/3 to
1.30. The model bounces at a relatively high density
($ \rho_c (t_{\rm bounce}) = 9.3 \times 10^{14} {\rm\ g\ cm}^{-3} $)
because the supranuclear EoS was chosen to be extremely soft
($ \gamma_{\rm p} (\rho > \rho_{\rm nuc}) = 5/3 $). After bounce the
rotating core oscillates, in both simulations, with a superposition of
various radial and axisymmetric modes. As expected, the oscillation
frequencies of the core are higher in the relativistic model due to
the higher densities achieved.

On the other hand, model~B is initially an almost rigid rotator, which
bounces later ($ t_{\rm bounce} = 98 {\rm\ ms} $) than model~A due to its
larger subnuclear adiabatic index $ \gamma_{\rm p} = 1.325 $. The collapse
is stopped at subnuclear densities
($ \rho_c (t_{\rm bounce}) = 3.8 \times 10^{13} {\rm\ g\ cm}^{-3} $)
by the rapid spin-up of the core due to angular momentum
conservation. After bounce the core first expands, then begins to
collapse once more and rebounces again. The phenomenon of homologous
multiple bounces~\cite{moenchmeyer_91} (which occurs with a period of
roughly 40~ms (50~ms) in the relativistic (Newtonian) simulation; see
lower panel in Fig.~\ref{fig:rho_c_t}) was observed in the Newtonian
simulations of Zwerger and M\"uller~\cite{zwerger_97} for strongly
differentially rotating configurations with a sufficiently large
initial value of $ \beta $ and a subnuclear $ \gamma_{\rm p} $ close
to $ 4/3 $. However, general relativistic models show distinct
multiple bounces only for extremely rapidly and differentially
rotating initial models. Most of the Newtonian models of Zwerger and
M\"uller, which bounce at subnuclear densities, reach or even exceed
in the relativistic simulations twice the maximum density of the
Newtonian models. Thus, in many cases the central density largely
exceeds nuclear matter density and the stiffening of the EoS destroys
the coherence of the bounce.

\begin{figure}[t]
  \def\epsfsize#1#2{0.48#1}
  \centerline{\epsfbox{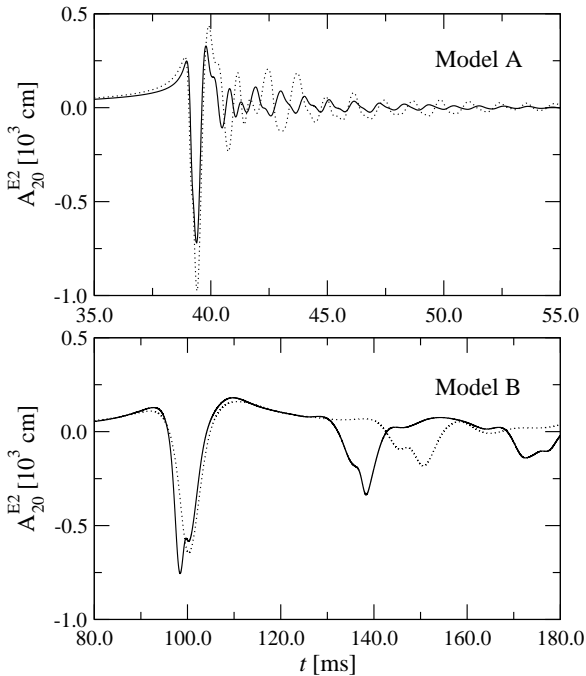}}
  \caption{Quadrupole gravitational wave amplitude for model~A (upper
    panel) and B (lower panel); relativistic runs are denoted by
    solid lines, Newtonian runs by dotted lines.}
  \label{fig:a_quad_t}
\end{figure}

The gravitational wave signature of both
models is shown in Fig.~\ref{fig:a_quad_t}, where we plot the
quadrupole wave amplitude $ A_{20}^{\rm E2} $ as defined
in~\cite{zwerger_97}. For a source at a distance $ R $ the
corresponding dimensionless strain is $ h^{\rm TT}_{\theta \theta} =
8.85 \times 10^{-24} \sin^2 \theta \cdot (A_{20}^{\rm E2} / 1000)
\cdot (10 {\rm\ Mpc} / R) $, where $ \theta $ is the angle between the
line of sight and the rotational axis of the core.

\begin{figure}[t]
  \def\epsfsize#1#2{0.48#1}
  \centerline{\epsfbox{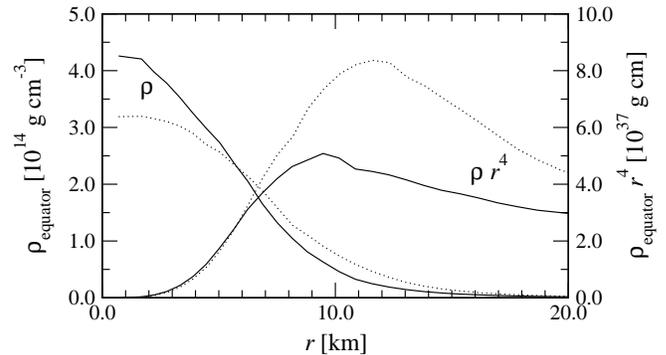}}
  \caption{Equatorial profiles of the density $ \rho $ and $ \rho r^4 $
    of model~A at bounce obtained in a Newtonian (dotted lines) and
    relativistic (solid lines) simulation.}
  \label{fig:rho_profile}
\end{figure}

For the rapidly collapsing model~A with its soft supranuclear EoS
the maximum central density of the relativistic simulation is almost 50\%
higher than in the Newtonian case. Nevertheless, the gravitational
wave amplitude of the relativistic model is lower than that of its
Newtonian counterpart. This surprising result can be understood by
noting that in the quadrupole formula the gravitational wave amplitude
is determined by the second time derivative of the total quadrupole
moment of the core, which involves a volume integral where the
integrand is proportional to $ \rho r^4 $. Thus a higher central
density due to relativistic effects does not necessarily imply a
stronger signal. Instead, one must take into account the whole density
distribution. In Fig.~\ref{fig:rho_profile} the profiles of $ \rho $
and $ \rho r^4 $ at the equator ($ \theta = \pi / 2 $) are plotted for
both the relativistic and the Newtonian model at the time of
bounce. The density profile of the relativistic model is more compact
than that of the Newtonian model, whereas the density of the latter
model is larger for radii $ \gsim 5 {\rm\ km} $~\cite{evans_85}. This
gives rise to its larger quadrupole moment (as the deformation of both
models is comparable), and hence to its larger gravitational wave
amplitude.

In model~B the relativistic effects cause the bounce to occur at
much higher densities, and the interval between the multiple bounces is
also shortened. In this case the gravitational wave amplitude of the
relativistic model is actually somewhat larger than in the Newtonian one.
As we have already pointed out, general relativistic effects may play
a key role in deciding whether a rapidly rotating core bounce will stall
at subnuclear density (with a very distinctive gravitational waveform
as in model~B) or reach nuclear density (with a waveform as in
model~A). Therefore, relativistic effects have to be taken into account
when modeling rotational core collapse. We will investigate this issue
further and present more quantitative results in an upcoming article.

As our relativistic simulations yield signal amplitudes similar to
Newtonian models, the prospects for detection of rotational core
collapse events are also similar. Therefore the predicted signal strength
of Galactic events lies within the sensitivity range of the upcoming
gravitational wave interferometers~\cite{thorne_97}. However,
extragalactic events seem to be too weak for current detector
sensitivities. To further improve the waveforms, 3-dimensional
effects and advanced microphysics have to be taken into account.

We appreciate contributions and suggestions from
Y.~Eriguchi, J.M$ ^{\underline{\mbox{\scriptsize a}}} $~Ib\'a\~nez, and
G.~Sch\"afer. The calculations were carried out at the Rechenzentrum
Garching and MPI f\"ur Astrophysik, Garching.

\end{document}